\documentclass[prl,twocolumn,showpacs,preprintnumbers,amsmath,amssymb]{revtex4}

 \def\be{\begin{equation}}
 \def\ee{\end{equation}}
 \def\bes{\begin{eqnarray}}
 \def\ees{\end{eqnarray}}
  \def\a{\alpha}
  \def\b{\beta}

 \def\G{\Gamma}

 \def\p{\partial}
 
 \def\t{\tau}

 \def\2{\frac{1}{2}}
 \def\4{\frac{1}{4}}

\catcode`\@=11
\def\@citex[#1]#2{%
\if@filesw \immediate \write \@auxout {\string \citation {#2}}\fi
\@tempcntb\m@ne \let\@h@ld\relax \def\@citea{}%
\@cite{%
  \@for \@citeb:=#2\do {%
    \@ifundefined {b@\@citeb}%
      {\@h@ld\@citea\@tempcntb\m@ne{\bf ?}%
      \@warning {Citation `\@citeb ' on page \thepage \space
undefined}}%
      {\@tempcnta\@tempcntb \advance\@tempcnta\@ne%
      \@tempcntb\number\csname b@\@citeb \endcsname \relax%
      \ifnum\@tempcnta=\@tempcntb 
it
        \ifx\@h@ld\relax%
          \edef \@h@ld{\@citea\csname b@\@citeb\endcsname}%
        \else%
          \edef\@h@ld{\ifmmode{-}\else--\fi\csname
b@\@citeb\endcsname}%
        \fi%
      \else
        \@h@ld\@citea\csname b@\@citeb \endcsname%
        \let\@h@ld\relax%
      \fi}%
    \def\@citea{,\penalty\@highpenalty\,}%
  }\@h@ld
}{#1}}

\def\@citeb#1#2{{[#1]\if@tempswa , #2\fi}}
\def\@citeu#1#2{{$^{#1}$\if@tempswa , #2\fi }}
\def\@citep#1#2{{#1\if@tempswa , #2\fi}}

\begin{document}
\preprint{UTHET-07-1101}

\title{Bjorken flow from an AdS Schwarzschild black hole}

\author{James Alsup}
\email{jalsup1@utk.edu}
\author{George Siopsis}
 \email{siopsis@tennessee.edu}
\affiliation{%
Department of Physics and Astronomy,
The University of Tennessee,
Knoxville, TN 37996 - 1200, USA.
}%
\date{May 2008}%

\begin{abstract}

We consider a large black hole in asymptotically AdS spacetime of arbitrary dimension with a Minkowski boundary.
By performing an appropriate slicing as we approach the boundary, we obtain via holographic renormalization a gauge theory fluid obeying Bjorken hydrodynamics
in the limit of large longitudinal proper time.
The metric we obtain reproduces to leading order the metric recently found as a direct solution of the Einstein equations in five dimensions.
Our results are also in agreement with recent exact results in three dimensions.

 \end{abstract}

\maketitle


{\em Introduction.--} Hydrodynamics has since the early eighties been used as a description for the strongly interacting particles produced in heavy ion collisions.  Such behavior has been supported by recent experiments performed at RHIC \cite{hydro}.  However, Quantum Chromodynamics (QCD) is marred by the difficulty of providing an explanation from first principles.  

The AdS/CFT correspondence~\cite{adscft,adscftrev} has provided another path for understanding such phenomena.  By studying a dual AdS$_d$ space one may gain insight into a strongly coupled gauge theory in one fewer dimension.  While the strongly coupled gauge field in five dimensions is a $\mathcal{N}=4$ super Yang-Mills theory, the results may be relevant to theories with less symmetry, such as QCD.  In the context of heavy ion collisions the dual description has been used to gain information about the strongly interacting plasma such as jet quenching and transport coefficients among others \cite{sinzahed,sonstarinets,natsuume,zahedsin,son}.

In order to understand a flowing hydrodynamic description of the gauge field one must introduce time dependence into the dual AdS space.  This was done in \cite{JP}, where dependence on the proper time in the longitudinal plane of the collision was introduced into an AdS$_5$ space.  Through the AdS/CFT correspondence it was found, in the late time limit, the boundary gauge field followed ideal Bjorken hydrodynamics \cite{Bjorken}.  The work has been furthered to understand the subleading terms in the expansion of the solution of the Einstein equations and the relation to dissipative hydrodynamics on the boundary \cite{NakSin,SinNakKim,ASM,Janik,BJ}.  

In an interesting recent work, Kajantie, Louko and Tahkokallio~\cite{KLT} found a time-dependent solution in three dimensions that also produced a Bjorken flow in the boundary gauge theory.  The time dependence of the solution could then be removed by a coordinate transformation to the standard AdS$_3$ Schwarzschild metric. Thus, a boost-invariant flow could be understood in terms of a static Schwarzschild black hole.
This is perhaps not surprising as three dimensions are special and results are not necessarily generalizable to higher dimensions.

Our aim is to show that the three-dimensional result of Kajantie, {\em et al.,} \cite{KLT} generalizes to arbitrary dimension.
By performing an appropriate slicing near the boundary, we shall obtain a Bjorken flow from a static Schwarzschild black hole via holographic renormalization \cite{Skenderis} to leading order in longitudinal proper time.
In three dimensions, our results reduce to those of ref.~\cite{KLT}.
In five dimensions, we recover the metric of Janik and Peschanski \cite{JP}.
Higher-order corrections can be calculated by a refinement of the slicing we perform here.


{\em Schwarzschild black hole.--} We start with a short discussion of pertinent properties of an AdS$_d$ Schwarzschild black hole.
Although generally known, we cast them in a form that facilitates application to the non-static case.
An AdS black hole is a solution of the Einstein equations
\be\label{eqEin}
R_{\mu\nu}-\left(\frac{1}{2}R+\Lambda_{d}\right)g_{\mu\nu}=0
\ee
where $\Lambda_{d} = -\frac{(d-1)(d-2)}{2}$ is a negative cosmological constant.
A large black hole has a flat horizon and may be found by substituting the ansatz
\be\label{eqmebh} ds_{\mathrm{b.h.}}^2 = \frac{1}{z^2} \left( -e^{a(z)} dt^2 + {d\vec x\,}^2 + e^{b(z)} dz^2 \right) \ee
where $\vec x \in \mathbb{R}^{d-2}$, in the Einstein equations.
They reduce to the two independent equations
\be\label{eq3} a' + b' = 0
\ \ , \ \ \ \ zb' + (d-1)(1-e^b) = 0 \ee
whose solution is
\be\label{eq4} a(z) = - b(z) = \ln \left( 1 - 2\mu z^{d-1} \right) \ee
where $\mu$ is an arbitrary integration constant.
The horizon is at
\be\label{eq5} z_+ = (2\mu)^{-\frac{1}{d-1}} \ee
and the boundary of the asymptotically AdS space is at $z=0$.

The Hawking temperature of the hole is
\be\label{eqTH} T_H = \frac{d-1}{4\pi z_+} \ee
This solution is related to a gauge theory on the boundary via 
holographic renormalization \cite{Skenderis}.  One may construct the vacuum expectation value of the gauge theory's stress-energy tensor from the form of the bulk metric.  The metric needs to be brought in the form of a general asymptotically AdS metric in Fefferman-Graham coordinates
\be\label{FG}
ds^2=\frac{g_{\mu\nu}dx^\mu dx^\nu+dz_{FG}^2}{z_{FG}^2}\ee
Then near the
boundary at $z_{FG}=0$ we may expand
\bes
g_{\mu\nu} &=& g_{\mu\nu}^{(0)}+z_{FG}^2 g_{\mu\nu}^{(2)} +
\dots+z_{FG}^{d-1}g_{\mu\nu}^{(d-1)}\nonumber\\
& & +h^{(d-1)}z_{FG}^{d-1}\ln z_{FG}^2 +\mathcal{O}(z_{FG}^d) \ees
where $g^{(0)}_{\mu\nu}=\eta_{\mu\nu}$.
In up to at least seven dimensions $g_{\mu\nu}^{(d-1)}$ is proportional to the vacuum expectation value of the stress-energy tensor,
\be\label{eqgT} \langle T_{\mu\nu}\rangle = \frac{d-1}{16\pi G_d} g_{\mu\nu}^{(d-1)}
\ee
where $G_d$ is Newton's constant in the bulk.

To find the stress-energy tensor corresponding to the hole, we write the radial distance in the bulk as
\be\label{eq10} z = z_{FG} \left[ 1 - \frac{\mu}{d-1} z_{FG}^{d-1} + \mathcal{O} (z_{FG}^{2(d-1)}) \right] \ee
so that the metric (\ref{eqmebh}) is of the Fefferman-Graham form.
We obtain
\be g_{tt}^{(d-1)} = \frac{d-2}{d-1} \, 2\mu \ \ , \ \ \ \ g_{ij}^{(d-1)} = \frac{2\mu}{d-1} \delta_{ij} \ee
($i,j = 1,\dots, d-2$)
leading to energy density and pressure of the gauge theory fluid on the boundary, respectively,
\be \varepsilon = \langle T^{tt} \rangle = (d-2) \frac{\mu}{8\pi G_d} \ \ , \ \ \ \ p = \langle T^{ii} \rangle = \frac{\mu}{8\pi G_d} \ee
obeying $p = \frac{1}{d-2} \varepsilon$, as expected for a conformal fluid.
With the temperature given by (\ref{eqTH}), we obtain the equation of state
\be p = \frac{1}{16\pi G_d} \left( \frac{4\pi T_H}{d-1} \right)^{d-1} \ee
and the energy and entropy densities, respectively, as functions of the temperature
\be\label{eqes} \varepsilon = \frac{d-2}{16\pi G_d} \left( \frac{4\pi T_H}{d-1} \right)^{d-1} \ \ , \ \ s = \frac{dp}{dT} = \frac{1}{4G_d} \left( \frac{4\pi T_H}{d-1} \right)^{d-2} \ee
%

{\em Bjorken Hydrodynamics.--} Having understood the case of a static gauge theory fluid on $(d-1)$-dimensional Minkowski space, we turn our attention to boost invariant hydrodynamics
in order to understand heavy ion collisions, following a suggestion by Bjorken \cite{Bjorken}.
The gauge theory fluid will still be on a $(d-1)$-dimensional Minkowski space as in the static case, but to distinguish it from
the boundary of the large AdS$_d$ Schwarzschild black hole, we shall denote its
coordinates by $\tilde x^\mu$ ($\mu = 0,1,\dots, d-2$) and assume that
the colliding beams are along the $\tilde x^1$ direction.
It is convenient to choose coordinates $\tau, y$ (proper time and rapidity in the longitudinal plane, respectively), where
\be
\tilde x^0=\tau \cosh y~~,~~~~\tilde x^1=\tau \sinh y
\ee
The $(d-1)$-dimensional Minkowski metric takes the form
\be\label{eq16}
d\tilde s^2=d\tilde x_\mu d\tilde x^\mu = -d\tau^2+\tau^2 dy^2+(d\tilde x^\bot)^2
\ee
where $\tilde x^\bot = (\tilde x^2,\dots,\tilde x^{d-2})$ represents the transverse coordinates.

For the stress-energy tensor let us assume one which satisfies boost invariance, symmetry under reflection in the longitudinal direction ($y\to -y$), plus translational and rotational invariance \cite{JP},
\be
T^{\mu\nu} =
\mathrm{diag} \left(
\ \varepsilon (\t)
\ \ p(\t)/\t^2 \ \ \dots
\ \ p(\t)
\ \right) \ .
\ee
Using the local conservation law for the stress-energy tensor,
\be\label{eq18}
\nabla_\a T^{\a\b}=\p_\a T^{\a\b}+\G^\a_{\a\lambda} T^{\lambda\b}+\G^\b_{\a\lambda} T^{\a\lambda} =0
\ee
with the Christoffel symbols $\G^y_{y\t}=\frac{1}{\t}=\G^y_{\t y}$ and $\G^\t_{yy}=\t$,
we derive relations between the components of the stress tensor.

Choosing $\b=\t$, we obtain
\be
\p_\t \varepsilon +\frac{1}{\t} (\varepsilon + p)  =0
\ee
Demanding tracelessness, a consequence of conformal invariance, we obtain another constraint on the stress-energy tensor
\be
-\varepsilon +(d-2)p=0
\ee
%
Solving the above equations, we deduce
\be\label{HydroEP}
\varepsilon = (d-2) p = \frac{\varepsilon_0}{\tau^{\frac{d-1}{d-2}}}
\ee
The temperature of the system may be found as a consequence of a perfect fluid's entropy conservation. We obtain \cite{Bjorken}
\be\label{HydroT}
T=\frac{T_0}{\tau^{1/(d-2)}}
\ee
The constants $\varepsilon_0$ and $T_0$ represent the initial values of the energy density and temperature, respectively (at $\tau = 1$).

The entropy density is
\be\label{eq23} s = \frac{\dot p}{\dot T} = \frac{s_0}{\tau} \ \ , \ \ \ \ s_0 = \frac{d-1}{d-2}\, \frac{\varepsilon_0}{T_0} \ee
Notice that the energy and entropy densities have the same dependence on the temperature as in the static case (eq.~(\ref{eqes})).
If we identify initial data with their corresponding values in the static case,
\be\label{eqTe} T_0 = T_H \ \ , \ \ \ \ \varepsilon_0 = \frac{(d-2)\mu}{8\pi G_d} \ee
then eq.~(\ref{eqes}), with $T_H$ replaced by $T$ (eq.~(\ref{HydroT})), describes the evolution of the energy and entropy densities in a Bjorken flow.

%
To find the solution of the Einstein equations (\ref{eqEin}) which follows the same symmetries as that of the stress-energy tensor, namely boost invariance, symmetry under reflection in the longitudinal direction ($y\to -y$), plus translational and rotational invariance, we shall adopt the ansatz
\be\label{metric}
ds^2_{\mathrm{Bjorken}} =\frac{-e^{A}d\tau^2+\tau^2 dy^2+e^{C} (d\tilde x^\bot)^2+e^{B} d\tilde z^2}{\tilde z^2}
\ee
where $A,B,C$ are all functions of $\tilde z$ and $\tau$, following \cite{JP,BJ}.
For the perfect fluid solution we also imposed the condition of isotropy $\frac{1}{\tau^2}g_{yy}=g_{ii}$. The coordinates need to be brought into the Fefferman-Graham form (\ref{FG}) so the hydrodynamics may be derived via holographic renormalization \cite{Skenderis}.

The Einstein equations will couple the dependence of $A,B,C$ on $z$ and $\tau$, but this problem is eliminated by introducing a variable $v$ which is kept fixed as $\t\to\infty$,
\be\label{eqv}
v=\frac{\tilde z}{\tau^{1/(d-2)}}
\ee
Assuming that the functions $A(\tilde z,\tau),~B(\tilde z,\tau),~C(\tilde z,\tau)$ become functions of only $v$,
\be\label{eq27} A = A_0(v) + \dots \ \ , \ \ \ \ B = B_0 (v) + \dots \ \ ,  \ \ \ \ C = C_0(v) + \dots \ee
where the dots represent terms that vanish in the $\tau\to\infty$ limit, the Einstein equations (\ref{eqEin}) are then reduced to the three independent equations
\be A_0' + B_0' = C_0' = 0
\ \ , \ \ \ \ v B_0' + (d-1)(1-e^{B_0}) = 0 \ee
which are of the same form as eq.~(\ref{eq3}) in the static case.
They are solved by
\be\label{bulkSoln}
A_0(v) = - B_0(v) = \ln \left( 1 - 2\mu v^{d-1} \right) \ \ , \ \ \ \ C_0(v) = 1 \ee
where, again, $\mu$ is an integration constant ({\em cf.}~with eq.~(\ref{eq4})).

In order to gain information of the gauge theory on the boundary, we use holographic renormalization \cite{Skenderis}.
The metric~(\ref{metric}) needs to be expressed in Fefferman-Graham coordinates (eq.~(\ref{FG})).
To leading order in $\tau$, this is achieved by the transformation
\be\label{eq30} \tilde z = \tilde z_{FG} \left[ 1 - \frac{\mu}{d-1} \frac{\tilde z_{FG}^{d-1}}{\tau^{(d-1)/(d-2)}} + \mathcal{O} (\tilde z_{FG}^{2(d-1)}) \right] \ee
which is similar to the static case (eq.~(\ref{eq10})).
For the metric (\ref{metric}) we may read off
\be\label{eq31}
\varepsilon = \langle T^{\tau\tau}\rangle = \frac{\varepsilon_0}{\tau^\frac{d-1}{d-2}}~~,~~~~p = \tau^{2} \langle T^{yy}\rangle=\langle T^{ii}\rangle = \frac{\varepsilon}{d-2}
\ee
where $\varepsilon_0$ is given by (\ref{eqTe}).
In comparison with (\ref{HydroEP}) we see that the geometry is the dual of Bjorken hydrodynamics.
However, the assignment of temperature and entropy is a little murky, because the bulk metric does not possess a static horizon.
The null surface at $v = (2\mu)^{-1/(d-1)} = z_+$ cannot be used for a rigorous definition of the temperature, because the bulk metric (\ref{metric}) with $A=A_0$ and $B=B_0$ (eq.~(\ref{bulkSoln})) is not an exact solution of the Einstein equations; it is only the leading term in a $1/\t$ expansion placing the null surface at the boundary of the region of validity of the expansion.
Nevertheless, if one blindly follows the arguments in the static case \cite{JP,BJ}, one obtains from (\ref{eqTH})
\be\label{eq32} T = \frac{d-1}{4\pi \tilde z_+} = \frac{T_H}{\tau^{1/(d-2)}} \ee
where we used (\ref{eqv}), in agreement with the result (\ref{HydroT}) from Bjorken hydrodynamics with initial data (\ref{eqTe}).
Knowing $T$, we may deduce the entropy density as in (\ref{eq23}).

Our goal now turns to understanding the bulk geometry in terms of a static AdS black hole and shed some light on the validity of the assignment of temperature (\ref{eq32}).


{\em Static to Flowing.--} In order to produce a flow on the boundary of the static black hole, instead of
approximating the boundary with $z=$~const.~hypersurfaces (as $z\to 0$), we shall make a different choice of slicing.

Near the boundary, the two metrics (\ref{eqmebh}) and (\ref{metric}) may be approximated respectively by
\bes\label{eq33}
ds_{\mathrm{b.h.}}^2 &\to& \frac{1}{z^2} \left( -dt^2+d{\vec{x}\,}^2+dz^2 \right) \nonumber\\
ds_{\mathrm{Bjorken}}^2 &\to& \frac{1}{\tilde z^2}\left(-d\tau^2+\tau^2 dy^2 + (d\tilde x^\bot)^2+d\tilde z^2\right)
\ees
While the former is the asymptotic form of an exact solution of the Einstein equations,
the latter is only valid in the large $\t$ limit.
We are interested in finding a transformation which relates the two asymptotic forms in this limit.
To be precise, we define the $\t\to\infty$ limit as follows: let
\be \tau = \tau_0 + \tau' \ee
where $\tau_0$ is a constant. We assume $\tau_0 \gg 1$ and $\tau' \sim \mathcal{O} (1)$ so that $d\tau = d\tau' \sim \mathcal{O} (1)$.
Also $\tilde x^\bot \sim \mathcal{O} (1)$ and $v\sim \mathcal{O} (1)$.
The latter implies $\tilde z \sim \mathcal{O} (\tau_0^{1/(d-2)})$.
By defining
\be \tilde z = \tilde z_0 \tau_0^{1/(d-2)} + \tilde z' \ee
and demanding $\tilde z_0, \tilde z' \sim \mathcal{O} (1)$, we ensure $d\tilde z = d\tilde z' \sim \mathcal{O} (1)$.
Of course, as we approach the boundary, we need to let both $\tilde z_0, \tilde z' \to 0$.
The remaining term in the metric will be $\mathcal{O} (1)$ provided we choose $y' \sim \mathcal{O} (1)$, where we defined
\be y = \frac{ y'}{\tau_0} \ee
Having thus defined the limit $\tau\to\infty$, it is not hard to see that the following transformation performs the desired task of relating the two metrics (\ref{eq33}),
\bes\label{eq34}
t=\frac{d-2}{d-3} \tau^{\frac{d-3}{d-2}}~~&,&~~~~x^1=\tau^{\frac{d-3}{d-2}} y \nonumber\\
x^\bot=\frac{\tilde x^\bot}{\tau^{1/(d-2)}}~~&,&~~~~z=\frac{\tilde z}{\tau^{1/(d-2)}}
\ees
Then, instead of the $z=$~const.~slicing, we shall approach the boundary on $\tilde z=$~const.~hypersurfaces (as $z,\tilde z \to 0$).
The latter coincide `initially' (at $\t = 1$), but ``flow'' as the new coordinates describing the black hole metric are $\tau$-dependent.


Applying the transformation (\ref{eq34}) to the exact black hole metric (\ref{eqmebh}) (more precisely, to a patch which includes the boundary $z\to 0$), we obtain
\bes\label{TransMetric}
ds^2_{\mathrm{b.h.}} &=& \frac{1}{\tilde z^2} \left[ - \left( 1-2\mu \frac{\tilde z^{d-1}}{\tau^{\frac{d-1}{d-2}}} \right) d\tau^2 + \tau^2 dy^2 + (d\tilde x^\bot)^2 \right. \nonumber\\
& & \left. + \frac{d\tilde z^2}{1-2\mu \frac{\tilde z^{d-1}}{\tau^{(d-1)/(d-2)}}} \right] + \mathcal{O} (\tau^{-1})
\ees
which matches the bulk metric of Bjorken flow (eqs.~(\ref{metric}), (\ref{eq27}) and (\ref{bulkSoln})) to leading order in $1/\t$.
Thus, the gauge theory fluid on the boundary of the Schwarzschild black hole which is approached with $\tilde z = $~const.~hypersurfaces as $\tilde z \to 0$ obeys Bjorken hydrodynamics in the large $\t$ limit.

In addition to the standard derivation of the energy density and pressure (\ref{eq31}), we may now address the issue of the temperature of the gauge theory fluid.
The horizon is static and the Hawking temperature is well-defined because the exact geometry giving rise to the approximate expression (\ref{TransMetric}) is a Schwarzschild black hole.
The Hawking temperature $T_H$ is the temperature of the static gauge theory fluid on the hypersurface $z\to 0$ whose metric is
\be\label{eqm1} ds_{z \to 0}^2 = - dt^2 + d\vec x^2 \ee
On the other hand, the $\tilde z \to 0$ hypersurface has metric
\be\label{eqm2} ds_{\tilde z \to 0}^2 = -d\t^2 + \t^2 dy^2 + (d\tilde x^\bot)^2 \ee
This Bjorken metric ({\em cf.}~with eq.~(\ref{eq16})) is related to the metric (\ref{eqm1}) in the large $\t$ limit by a conformal transformation which is obtain by restricting the transformation (\ref{eq34}) to these hypersurfaces,
\be\label{eq34b}
t=\frac{d-2}{d-3} \tau^{\frac{d-3}{d-2}}~~,~~~~x^1=\tau^{\frac{d-3}{d-2}} y ~~,~~~~
x^\bot=\frac{\tilde x^\bot}{\tau^{1/(d-2)}}
\ee
The two metrics (\ref{eqm1}) and (\ref{eqm2}) are related by
\be
ds_{z \to 0}^2 = \tau^{-\frac{2}{d-2}}\left[ ds_{\tilde z \to 0}^2 + \mathcal{O}(1/\tau) \right]
\ee
showing that the Euclidean proper time period of thermal Green functions
on the Bjorken boundary (\ref{eqm2}) scales as $\tau^{1/(d-2)}$.
Since the period is inversely proportional to the temperature, the latter scales as $\tau^{-1/(d-2)}$, in agreement with expectations (eq.~(\ref{eq32})).
The two hypersurfaces coincide at $\t = 1$ at which time $T=T_H$.

Let us also check that the transformation (\ref{eq34}) reduces to the transformation found in \cite{KLT} for $d=3$.  To do this we must proceed with a little care.  For the time coordinate we must add an appropriate constant term so that the limit $d\to3$ is well-defined.  Then as $d\to 3$, we obtain
\be
t=\ln~\tau~~,~~~~x^1 =y~~,~~~~z=\frac{\tilde z}{\tau}
\ee
The transformation to Fefferman-Graham coordinates can be found exactly in this case,
\be z = \frac{\tilde z_{FG}}{\tau} \left( 1 + \frac{\mu}{2}\, \frac{\tilde z_{FG}^2}{\tau^2} \right)^{-1} \ee
which matches the result of \cite{KLT} in the large black hole limit.
It also agrees with the general expression (\ref{eq30}) to first order.

Higher-order corrections to Bjorken flow dictated by the black hole may be found by refining the transformation (\ref{eq34}). This entails introducing corrections which are of $o(1/\tau)$ and making sure that the application of the transformation to the metric (\ref{eqmebh}) does not introduce dependence of the metric on the rapidity and the transverse coordinates.
This can be done systematically at each order in the $1/\tau$ expansion and will be reported on elsewhere.


{\em Conclusion.--} We discussed the possibility of obtaining Bjorken hydrodynamics \cite{Bjorken} on a $(d-1)$-dimensional Minkowski space from a large AdS$_d$ Schwarzschild black hole (of flat horizon).
The latter is normally considered dual to a static gauge theory fluid on the boundary whose temperature coincides with the Hawking temperature.
By introducing an appropriate set of coordinates in a patch of the hole which included the boundary, we obtained a generalization of the metric of Janik and Peschanski~\cite{JP} to arbitrary dimensions in the late time limit.
Thus, we obtained Bjorken hydrodynamics on the boundary in the limit of longitudinal proper time $\t \to\infty$.
This was effectively achieved by a slicing near the boundary of the black hole consisting of ``flowing'' hypersurfaces related to the standard static hypersurfaces by a time-dependent conformal transformation.
The conformal factor also provided a justification for determining the temperature.
Our results coincided with those of ref.~\cite{KLT} in the large black hole limit in three dimensions.

It would be interesting to see if the AdS$_d$ Schwarzschild metric (or other exact solutions of the Einstein equations) may be used to study subleading terms in the $\tau$ expansion of time-dependent solutions of the Einstein equations thereby encoding the effects of viscosity or heat conduction of the gauge theory fluid on the boundary \cite{NakSin,SinNakKim,ASM,Janik,BJ}.  Work in this direction is in progress.

\section*{Acknowledgment}
Work supported in part by the Department of Energy under grant DE-FG05-91ER40627.

\end{document}